\documentclass[a4paper,twocolumn,superscriptaddress]{revtex4}

\usepackage{graphicx}

\begin{document}

\title{Attosecond Control of Ionization Dynamics}

\author{P. Johnsson}
\email{per@eng-johnsson.se}
\affiliation{Department of Physics, Lund
University, P. O. Box 118, SE-221 00 Lund, Sweden}

\author{J. Mauritsson}
\affiliation{Department of Physics, Lund University, P. O. Box 118,
SE-221 00 Lund, Sweden}
\affiliation{Department of Physics and
Astronomy, Louisiana State University, Baton Rouge, Louisiana
70803-4001, USA}

\author{T. Remetter}
\affiliation{Department of Physics, Lund University, P. O. Box 118,
SE-221 00 Lund, Sweden}

\author{A. L'Huillier}
\affiliation{Department of Physics, Lund University, P. O. Box 118,
SE-221 00 Lund, Sweden}

\author{K. J. Schafer}
\affiliation{Department of Physics and Astronomy, Louisiana State
University, Baton Rouge, Louisiana 70803-4001, USA}

\date{\today}

%%%%%%%%%%%%%%%%%%%%%%%%%%%%%%%%%%%%%%%%%%%%%%%%%%%%%%%%%%%%%%%%
%%%%%%%%%%%%%%%%%%%%%%%%%%%%%%%%%%%%%%%%%%%%%%%%%%%%%%%%%%%%%%%%
% Introductory paragraph
\begin{abstract}
Attosecond pulses \cite{PaulScience2001, DrescherScience2001} can be
used to initiate and control electron dynamics on a sub-femtosecond
time scale. The first step in this process occurs when an atom
absorbs an ultraviolet photon leading to the formation of an
attosecond electron wave packet (EWP). Until now, attosecond pulses
have been used to create free EWPs in the continuum, where they
quickly disperse~\cite{KienbergerScience2002, KienbergerNature2004,
GoulielmakisScience2004, JohnssonPRL2005, RemetterNP2006}. In this
paper we use a train of attosecond pulses, synchronized to an
infrared (IR) laser field, to create a series of EWPs that are below
the ionization threshold in helium.  We show that the ionization
probability then becomes a function of the delay between the IR and
attosecond fields. Calculations that reproduce the experimental
results demonstrate that this ionization control results from
interference between transiently bound EWPs created by different
pulses in the train. In this way, we are able to observe, for the
first time, wave packet interference in a strongly driven atomic
system.
\end{abstract}

\maketitle

The modulation of photon absorption by wave packet interference
(WPI) has been used in molecular systems as a probe of nuclear
dynamics on a femtosecond time scale~\cite{SchererJCP1991}, and in
Rydberg atoms as a probe of electron dynamics on a picosecond time
scale~\cite{NoordamPRA1992}. WPI is a sensitive tool for probing
quantum dynamics because it depends on the spatial and temporal
behavior of the wave packet in the confining potential. This is most
easily appreciated by considering two well separated excitation
pulses that create two initially localized wave packets in a
manifold of excited states. The total excitation probability will be
the simple sum of the integrated population in each wave packet {\it
unless} some part of the first wave packet returns to the region of
space where it was created during the time when the second wave
packet is created, enabling the two to interfere.

\begin{figure}[tbh]
\includegraphics[width=1.0\linewidth]{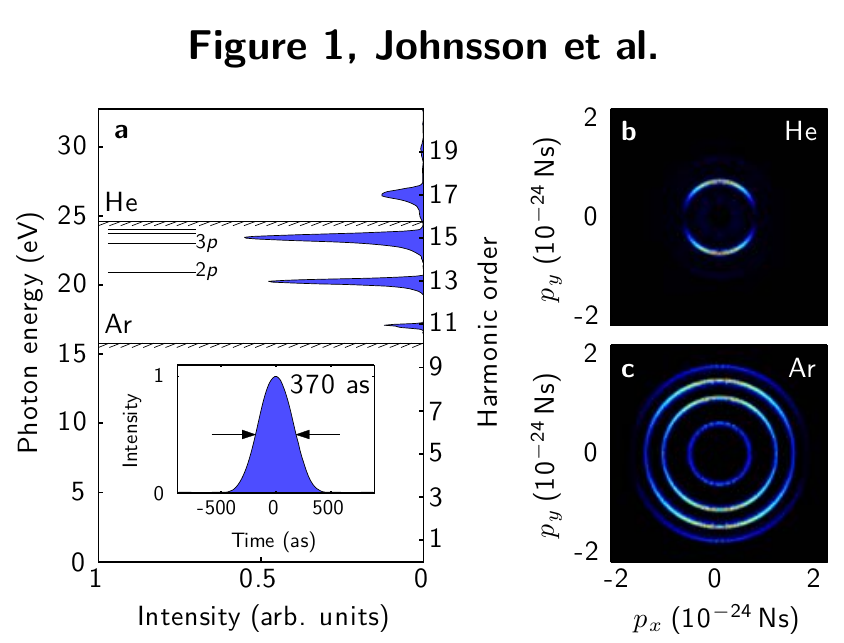}
\caption{\label{fig:train} \textbf{Experimental pulses and
single-photon momentum distributions.} \textbf{a}, Spectrum of the
UV pulses used in the experiment shown in relation to the ionization
potentials of helium and argon. For helium, some of the excited
states have been indicated for comparison. The corresponding
temporal profile is a train of pulses spaced by half the IR laser
cycle. The inset shows the temporal profile of the attosecond pulses
in the train, each with a duration of 370~as, as reconstructed from
the RABITT measurements. Panels \textbf{b} and \textbf{c} show
experimental photoelectron momentum distributions from single-photon
ionization by the APT in helium and argon, respectively, with the
polarization of the light parallel to the $p_y$-axis. \textbf{b}, In
helium, only a single ring corresponding to ionization by harmonic
17 can be seen, since this is the only spectral component having
sufficient energy to overcome the ionization potential. The angular
distribution is peaked along the polarization axis, as expected for
single-photon ionization from an s-state. \textbf{c}, In argon, the
full bandwidth of the APT contributes to the ionization and four
rings corresponding to harmonics 11 to 17 can be seen. Since the
ionization starts from a \emph{p}-state, the resulting angular
distribution is a superposition of \emph{s}- and \emph{d}-states,
with contributions also along the $p_x$-axis.}
\end{figure}

In this Letter we present attosecond WPI experiments using a train
of ultraviolet (UV) attosecond pulses to ionize either helium or
argon atoms in the presence of an IR field. The attosecond pulses
are phase locked to the IR field since their spacing in time is
precisely one half of the laser period. The central energy of the
pulses, $\approx$23~eV, is higher than the ionization energy of
argon (15.8~eV), but below that of helium (24.6~eV), as shown in
Fig.~\ref{fig:train}a. We demonstrate the ability to control the ion
yield in helium through the delay between the two fields, an effect
which is absent in argon. We attribute this ionization control to
interference between transiently bound EWPs created in helium which
can modulate the probability that an electron is excited out of the
atomic ground state. Calculations based on integration of the
time-dependent Schr\"odinger equation (TDSE) show that the contrast
in the ionization probability versus the IR-UV delay is an order or
magnitude larger that what is achieved with a single pulse, and that
the contrast grows as the number of pulses in the train is
increased.  Both of these effects are hallmarks of WPI, seen here in
the attosecond domain and for a strongly driven system.

%%%%%%%%%%%%%%%%%%%%%%%%%%%%%%%%%%%%%%%%%%%%%%%%%%%%%%%%%%%%%%%%%%%%%%%%%
% Method summary
Details of the experiment can be found in the Methods section. The
spectral and temporal characteristics of the attosecond pulse train
(APT) used to excite the atoms are presented in
Fig.~\ref{fig:train}a, while examples of momentum distributions
obtained from UV ionization alone are shown in Fig.~\ref{fig:train}b
and \ref{fig:train}c for helium and argon, respectively. The IR
laser field, a replica of the laser pulse used to generate the UV
pulses, was recombined with the APT after a variable delay line and
focused into the detection chamber with an intensity of
$1.3\times10^{13}$~W$\cdot$cm$^{-2}$. A crucial point is that this
laser intensity is too low to excite any population out of the
ground state by itself. This means that the ground state is
connected to the excited bound and continuum states only when an
attosecond pulse is present, an essential condition for observing
WPI. Also of importance is the fact that although the IR laser field
is weak from the point of view of an electron in the ground state,
it is strong from the point of view of an electron excited out of
the ground state. At peak amplitude, the IR field suppresses the
Coulomb potential by $\sim$7~eV at the saddle point, which is enough
to unbind all of the single excited bound states of helium.
Furthermore, this barrier suppression changes very slowly with
intensity, scaling as $I^{1/4}$. Our method therefore results in
creating attosecond EWPs in a strong oscillating laser field.

\begin{figure}[tbh]
\includegraphics[width=1.0\linewidth]{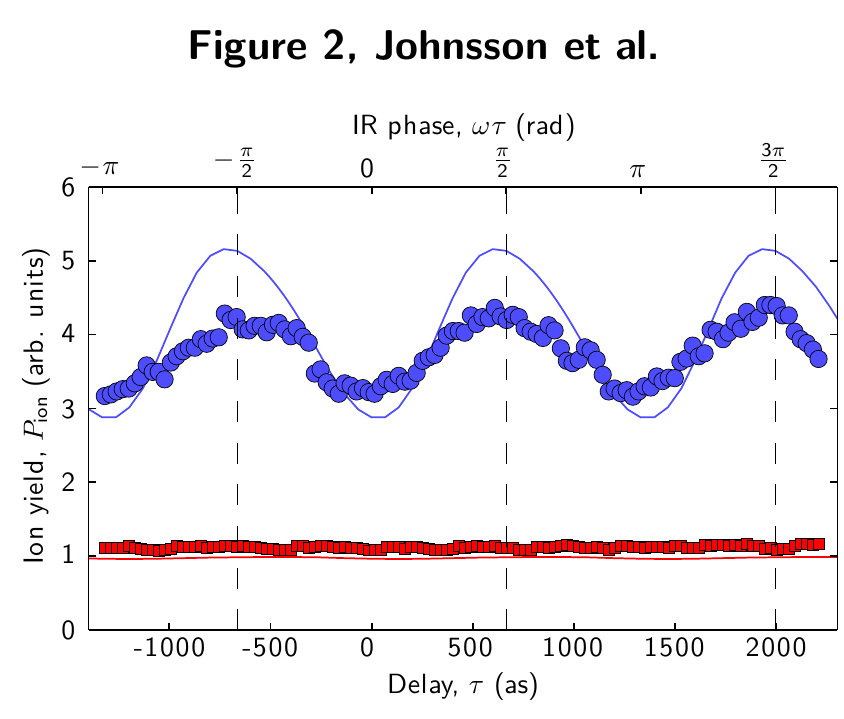}
\caption{\label{fig:ions} \textbf{Control of the ion yield.}
Experimentally measured ion yields, $P_{\mathrm{ion}}$, for He$^+$
(blue circles) and Ar$^+$ (red squares) as a function of the delay
$\tau$ between the attosecond pulses and the IR field, at an
intensity of $1.3\times10^{13}$~W$\cdot$cm$^{-2}$. We use a sine
convention for the IR electric field so that delays, $\tau$, which
are multiples of $\pi/\omega=1330$~as, where $\omega$ is the IR
laser frequency ($\hbar\omega=1.56$~eV), correspond to the
attosecond pulses overlapping the zero-crossings of the IR field.
All yields are normalised to those obtained with only the APT
present. In helium a clear modulation is observed which is not seen
in argon. Also shown in the figure are the calculated ion yields at
an IR intensity of $1.3\times10^{13}$~W$\cdot$cm$^{-2}$ for He$^+$
(blue solid line) and Ar$^+$ (red solid line). These were obtained
using UV and IR fields that closely match the experimental
parameters: the APT has a 10 fs FWHM (full width at half maximum in
intensity) duration and the IR pulse is 35 fs FWHM, and agree well
with the experimental results.}
\end{figure}

% Ions experiment
Figure~\ref{fig:ions} shows our main experimental result, the delay
dependence of the ion yields, $P_{\mathrm{ion}}$, from helium and
argon. For Ar$^+$ (red squares), there is no measurable effect of
the IR field while for He$^+$ (blue circles) the ion yield is
increased by a factor of four when the IR field is present. In
addition, the He$^+$ yield exhibits a modulation as a function of
the UV-IR delay. The depth of the modulation is $\approx35\%$ and
the period is equal to half the laser period. This modulation is the
signature of WPI of attosecond EWPs in our experiment.

% Ions theory
To gain insight into the results presented in Fig.~\ref{fig:ions},
we have performed calculations based on the integration of the
TDSE~\cite{KulanderAILF1992,KrausePRL1992}, as explained in more
detail in the Methods section. The ion yields obtained at the end of
the interaction are indicated in Fig.~\ref{fig:ions} as solid red
and solid blue lines for argon and helium respectively, showing good
agreement with the experiment. In addition, the calculations show
that without an IR field, the ionization probability in He is equal
to the excitation probability, meaning that no population is left in
the excited bound states. As indicated in Fig.~\ref{fig:train}a, the
spectrum of the UV pulses overlaps poorly with the accessible
excited bound states of helium, and the atom is limited to absorbing
photons belonging to the 17th harmonic, leading to immediate
ionization.

\begin{figure}[tbh]
\includegraphics[width=1.0\linewidth]{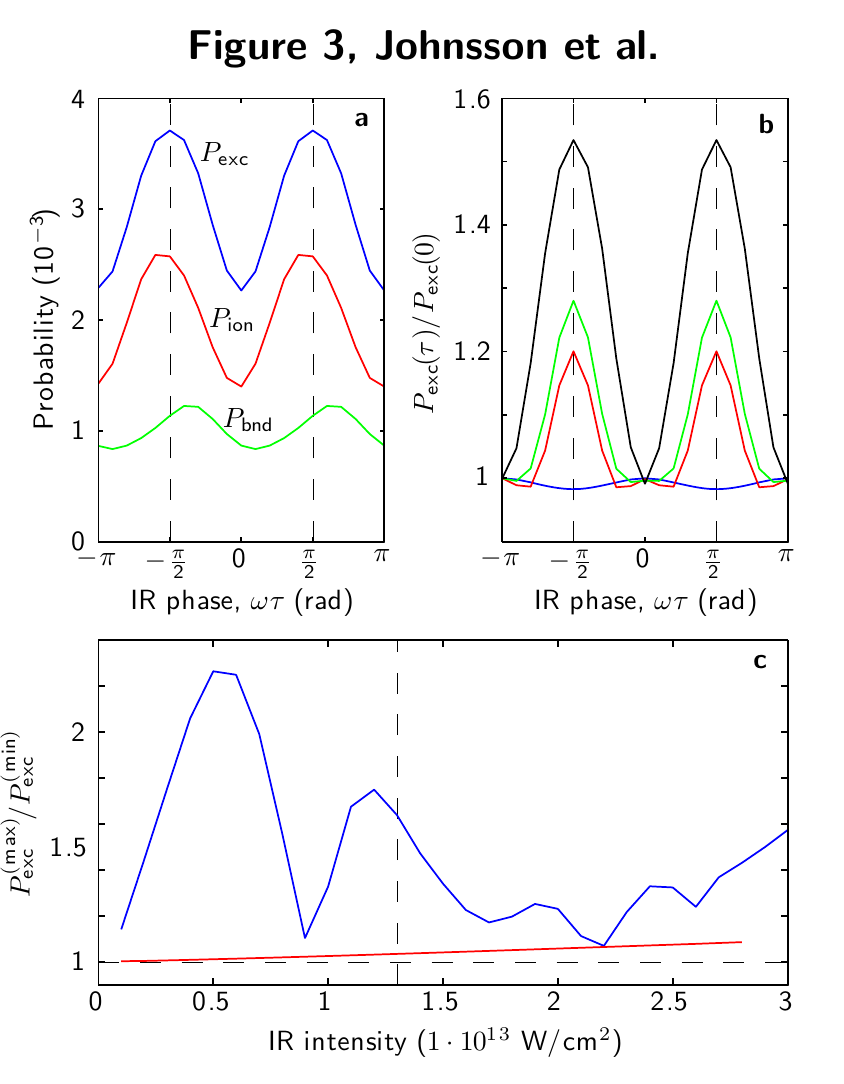}
\caption{\label{fig:ions_theory} \textbf{Detailed theoretical study
of ionization in helium.} \textbf{a}, Calculated probabilities for
removal of an electron from the ground state ($P_{\mathrm{exc}}$,
blue line), ionization ($P_{\mathrm{ion}}$, red line) and remaining
in an excited bound state ($P_{\mathrm{bnd}}$, green line) as a
function of the phase of the IR field at the time of the attosecond
pulses, for an APT with a FWHM of 10~fs and a 35~fs IR field with a
peak intensity of $1.3\times10^{13}$~W$\cdot$cm$^{-2}$. \textbf{b},
Excitation probability, $P_{\mathrm{exc}}$, versus delay for
different APTs, normalized to the excitation probability for zero
delay in each case. The FWHM of the APT intensity envelope is 1~fs
(blue line), 2~fs (red line), 4~fs (green line) or 8~fs (black
line). The 1~fs envelope corresponds to an isolated attosecond
pulse. \textbf{c}, Contrast in $P_{\mathrm{exc}}$ (defined as the
maximum excitation probability divided by the minimum) for various
peak intensities of the IR field (blue line). The APT has a FWHM of
10~fs. For comparison, the contrast obtained with a single
attosecond pulse is also shown (red line).}
\end{figure}

In Fig.~\ref{fig:ions_theory}a, we show more complete theoretical
results for He. Shown are both the probability that an electron is
excited out of the ground state ($P_{\mathrm{exc}}$, blue line) and
the ionization probability ($P_{\mathrm{ion}}$, red line). The
difference between these probabilities is the probability to remain
in an excited state after the IR pulse ends ($P_{\mathrm{bnd}}$,
green line). Two features are immediately apparent. First, the
modulation in the He$^+$ yield is caused by the fact that the amount
of population excited out of the ground state by the APT in the
presence of the IR field is modulated as a function of UV-IR delay.
Second, the ionization of the population promoted out of the ground
state by the APT is incomplete, leaving 30-40\% of the promoted
population in excited states after the IR field is over.

% Excitation step
The delay dependence of the He$^+$ yield has two contributions.
First, each pulse in the APT excites population in the presence of
an IR field that distorts the atomic potential by an amount that
depends on the IR-UV delay. A single attosecond pulse would
therefore probe solely the atom's ability to absorb light near the
ionization threshold in the presence of an electric field which can
be as high as  $\sim 10^8$~V/cm. Our calculations show that the
modulation in the ion yield due to such a single attosecond pulse is
about 1-3\% over the intensity range covered by the experiment, 10
times smaller than the observed effect. The other contribution to
the delay dependence is from WPI. This temporal interference in the
total excitation probability comes about if an EWP created by one
pulse in the train has some probability to be near the ion core when
a later packet is being excited by a different pulse in the train.
This requires that an EWP excited by a single pulse takes more that
one half cycle to completely ionize. Indeed, at all delays we find
that the EWP excited by a single attosecond pulse takes one to
several IR cycles to completely ionize, fulfilling this condition
for WPI.

WPI also causes the excitation probability to scale non-linearly
with the number of pulses in the train. In the absence of WPI the
relative modulation in the total excitation probability versus delay
is the same for different length pulse trains. In
Fig.~\ref{fig:ions_theory}b, we plot the normalized excitation
probability for APTs of different length, changing from a single
pulse (the 1 fs envelope) to two or more. We see that the relative
modulation increases as the APT length is increased. We also note
that the delay curve reverses its shape when the number of pulses is
increased from one to two or more. In argon, by contrast, the total
excitation is linear in the length of the pulse train.

By its nature, WPI is a very sensitive probe of the electron
dynamics in a bound system. In our system these dynamics are most
easily altered by changing the IR intensity. In
Fig.~\ref{fig:ions_theory}c we plot the magnitude of the calculated
relative modulation (\emph{i.e.} the contrast) versus peak IR
intensity for intensities ranging from 0.1 to $3.0
\times10^{13}$~W$\cdot$cm$^{-2}$ and a 10 fs APT (blue line). As can
be seen, the  contrast is a very sensitive function of the field
amplitude. For comparison, the contrast from using a single 370~as
pulse is shown (red line). In this same range of intensities the
amount of population ionized after the IR pulse is over ranges from
40-100\% of the total population excited out of the ground state,
and exhibits a very complicated dependence on the IR intensity.

\begin{figure}[tbh]
\includegraphics[width=1.0\linewidth]{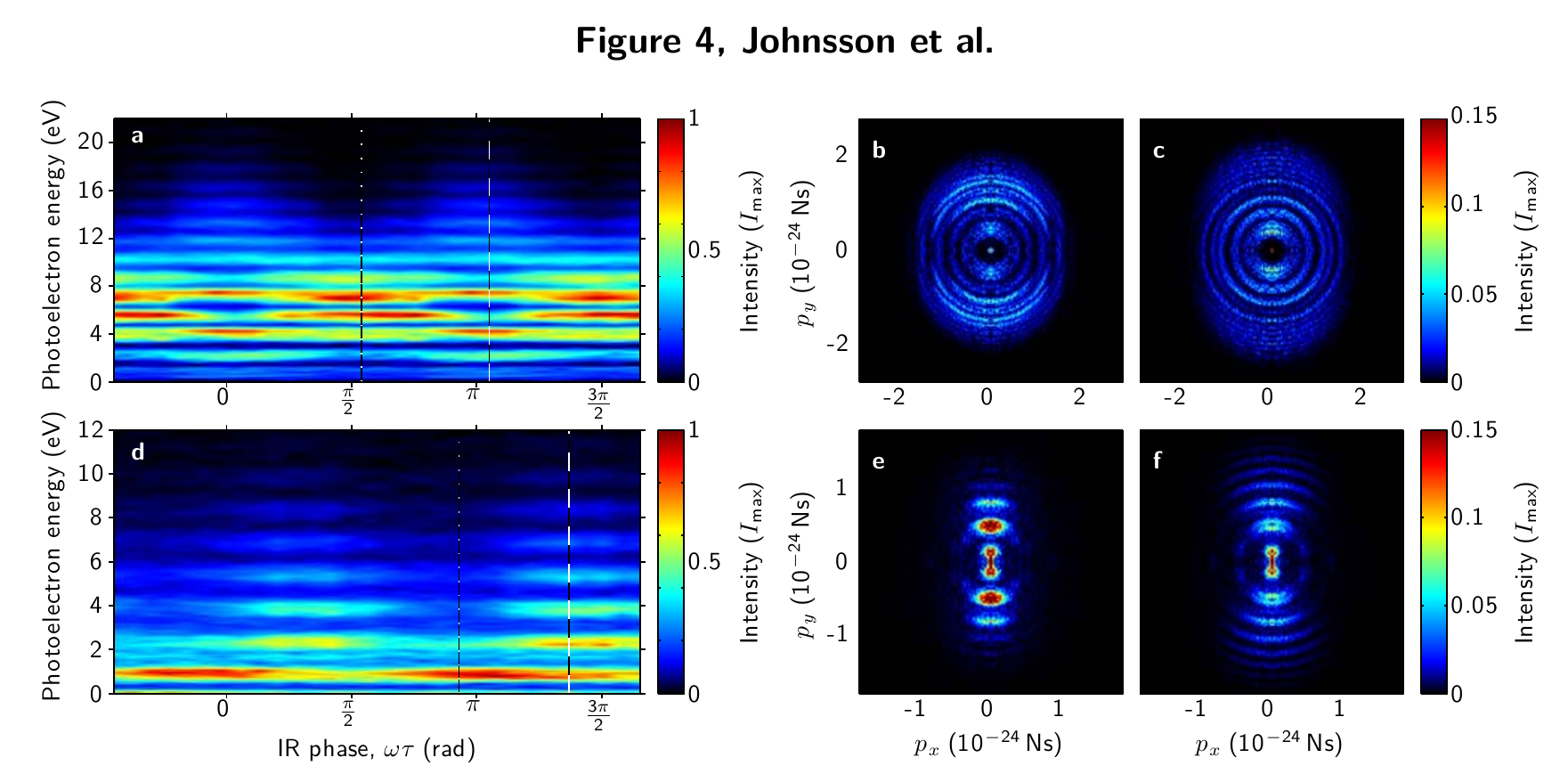}
\caption{\label{fig:electrons} \textbf{Experimental photoelectron
spectra and momentum distributions.} Photoelectron spectra as a
function of the delay between the attosecond pulses and the IR field
from argon (\textbf{a}) and helium (\textbf{d}). \textbf{b} and
\textbf{c}, Momentum distributions obtained at the delays
corresponding to the dotted and dashed lines in \textbf{a}.
\textbf{e} and \textbf{f}, Momentum distributions obtained at the
delays corresponding to the dotted and dashed lines in \textbf{d}.
The polarization directions of the UV and IR fields are parallel to
the $p_y$-axis.}
\end{figure}

% Electron results
Additional support for the WPI picture that we present can be found
in the experimental measurements of the energy-resolved angular
distributions from helium and argon, presented in
Fig.~\ref{fig:electrons}. For argon (Fig.~\ref{fig:electrons}a) the
IR field only redistributes the energy of the ionized electrons,
depending on the phase of the IR field at the time they enter the
continuum. The highest energy electrons are created when the
attosecond pulses are timed so that ionization takes place at the
zero-crossings of the electric field ($\omega\tau=n\pi$), when the
momentum transfer from the field to the electronic wave packet is
maximum~\cite{GoulielmakisScience2004, JohnssonPRL2005}. The
momentum distributions from argon (Fig.~\ref{fig:electrons}b and
\ref{fig:electrons}c) show the difference between the two delays
that results in the greatest and least number of high energy
electrons. The angular distributions remain rather broad for all
delays. In contrast to this, the photoelectron momentum
distributions from helium (Fig.~\ref{fig:electrons}e and
\ref{fig:electrons}f) are strongly peaked along the polarization
axis of the IR field, reflecting the fact that most of the
ionization occurs via electrons that escape over the suppressed
Coulomb barrier along the polarization direction. In addition, at
the experimental IR intensity the highest energy electrons are
observed when $\omega\tau \approx 0.3\pi + n\pi$, which corresponds
neither to the maxima or zeros of the IR electric field, as seen in
Fig.~\ref{fig:electrons}d. This illustrates the complex wave packet
dynamics discussed previously, which leads to different ratios
between bound and free populations as well as between numbers of low
and high energy electrons depending on the UV-IR delay and the IR
intensity.

%%%%%%%%%%%%%%%%%%%%%%%%%%%%%%%%%%%%%%%%%%%%%%%%%%%%%%%%%%%%%%%%%%%%%%%%%
%Discussion
The WPI that we have observed has a number of similarities and a few
important differences as compared to ``traditional'' WPI. In more
conventional WPI, the motion takes place on a purely bound potential
surface and the WPI is controlled by changing the delay between
pulses.  In our case, the delay between attosecond pulses is fixed
at one half the IR cycle, but the amplitude and phase of the IR
field  at which the EWPs are created is easily changed. Also, the
EWPs are only transiently bound and so both the total population and
the energy-resolved angular distributions can be measured as a
function of the various parameters in the experiment and compared to
theory. WPI offers a unique tool to study the behavior of electrons
in a strongly driven atom or molecule, since the EWPs are created in
the center of the potential well at a well-controlled time.

A number of modifications to the experiments we have presented here
are accessible in the near future. For instance, the wavelength of
the laser field can be varied, perhaps all the way to the
mid-infrared, which would allow the time difference between the
attosecond pulses to be varied. Most importantly perhaps, it should
be possible to study the WPI as a function of the APT duration as
was done in the theoretical calculations. This could be done in a
polarization gating scheme by varying the gate
duration~\cite{SolaNP2006}.

In conclusion, we have demonstrated that excitation/ionization
dynamics can be controlled using an APT in combination with an IR
field. Previous attosecond experiments have used the UV pulse to
control the time at which an ionization process takes
place~\cite{KienbergerScience2002, KienbergerNature2004,
DrescherNature2002, GoulielmakisScience2004, JohnssonPRL2005,
RemetterNP2006}. The control demonstrated in this experiment is, to
the best of our knowledge, the first use of an attosecond pulse to
modulate the probability of an ionization event. When coupled to
angular-resolved photoelectron distributions it opens the way for
future study of the detailed dynamics of ultra broadband EWPs in
driven atomic and molecular systems.

%%%%%%%%%%%%%%%%%%%%%%%%%%%%%%%%%%%%%%%%%%%%%%%%%%%%%%%%%%%%%%%%
%%%%%%%%%%%%%%%%%%%%%%%%%%%%%%%%%%%%%%%%%%%%%%%%%%%%%%%%%%%%%%%%
% Methods
\section*{Methods}
\subsubsection*{Experiment}
The APT was synthesized from high-order harmonics generated in xenon
by focusing 35~fs, 796~nm (1.56~eV photon energy) pulses from a
1~kHz Ti:sapphire laser to an intensity of
$\approx8\times10^{13}$~W$\cdot$cm$^{-2}$ in a 3~mm long windowless
gas cell filled to a static pressure of $\approx20$~mbar. The APT
was filtered spatially by passing it through a 1.5~mm diameter
aperture, and spectrally using a 200~nm thick aluminium filter. The
spatial filter removes contributions to the harmonic emission from
the longer quantum paths, while the aluminium filter blocks the
remaining IR and the intense low-order
harmonics~\cite{LopezMartensPRL2005}. The spectrum of the APT is
shown in Fig.~\ref{fig:train}a and consists of harmonics 11 to 17,
with a central energy of 23~eV. The pulses were characterised using
the RABITT technique (reconstruction of attosecond beating by
interference of two-photon transitions)~\cite{MullerAPB2002,
PaulScience2001}, and the average duration of the bursts was found
to be 370~as with the temporal profile shown in the inset in
Fig.~\ref{fig:train}a.

The dressing IR pulse, a delayed replica of the pulse generating the
harmonics, was collinearly overlapped with the APT before both beams
were refocused into the spectrometer by a toroidal platinum mirror.
Recombination was achieved using a mirror with a hole in the centre,
through which the APT was sent, and on which the dressing pulse was
reflected. The delay of the dressing pulse relative to the APT was
finely controlled by a mirror mounted on a piezo-electric
translation stage. To obtain a good estimate of the IR intensity,
calibration was performed by measuring Xe$^+$ and Xe$^{2+}$ ion
yields as a function of intensity. In addition, these estimates were
confirmed by comparison with the observed ponderomotive
shift~\cite{BucksbaumJOSAB1987} in the photoelectron spectra. The
absolute timing between the APT and the IR field was not accessible
experimentally, and has been chosen to fit the results of the TDSE
calculations.

A velocity map imaging spectrometer was used for detection, having
the advantage of being able to operate either in electron
imaging~\cite{EppinkRSI1997} or in ion time-of-flight mode. For both
ion and electron detection, the target gas was injected by means of
an atomic beam pulsed at 50~Hz. The ion yield measurements were
carried out in time-of-flight mode, the signal being collected with
a boxcar integrator using active background subtraction based on the
laser shots arriving with no target gas present. The 2D projections
of the momentum distributions of the photoelectrons were recorded by
means of an MCP-assembly and a CCD-camera, and from these the 3D
momentum distributions were obtained using the iterative inversion
procedure described in~\cite{VrakkingRSI2001}.

\subsubsection*{Theory}
For the TDSE calculations we use  the single active electron (SAE)
approximation, in which we assume that only one electron interacts
with the field, while the others remain in the ground state. This
approximation has been extensively tested for both alkali metal and
rare gas atoms and found to produce results which compare well with
experiments \cite{WalkerPRL1994,GaardePRL2000}. The atomic
potentials used in the SAE calculations were the standard
Hartree-Fock potential for helium, and a pseudo-potential in argon
\cite{KulanderCPC1991}. These potentials reproduce the single
electron excited states very well. To simulate the experiments we
use an IR pulse whose electric field envelope is a cosine function
with a FWHM in intensity of 35 fs, and an APT whose electric field
envelope is a somewhat sharper $\cos^2$ function with a FWHM in
intensity of 10~fs. The total population excited is calculated as
one minus the population remaining in the ground state at the end of
the pulse, while the total ionization is calculated either from the
photoelectron spectrum \cite{SchaferPRA1990} or by running the
calculation for 10 additional IR cycles and calculating the
probability to remain in the vicinity of the ion. A variety of other
field envelopes (\emph{e.g.} APTs with varying numbers of pulses and
a constant IR pulse envelope) were used to check details of our
analysis.

%%%%%%%%%%%%%%%%%%%%%%%%%%%%%%%%%%%%%%%%%%%%%%%%%%%%%%%%%%%%%%%%%%%%%%%%%
%%%%%%%%%%%%%%%%%%%%%%%%%%%%%%%%%%%%%%%%%%%%%%%%%%%%%%%%%%%%%%%%%%%%%%%%%
% Acknowledgements
\section*{Acknowledgments}
We are grateful to S.~Thorin, Dr. F.~L\'epine and Prof.
M.~J.~J.~Vrakking for help with the imaging spectrometer. This
research was supported by the Marie Curie Research Training Network
(MRTNCT-2003-505138, XTRA), the Crafoord Foundation, the Knut and
Alice Wallenberg Foundation, the Swedish Research Council and the
National Science Foundation through Grant No. PHY-0701372.

%%%%%%%%%%%%%%%%%%%%%%%%%%%%%%%%%%%%%%%%%%%%%%%%%%%%%%%%%%%%%%%%%%%%%%%%%
%%%%%%%%%%%%%%%%%%%%%%%%%%%%%%%%%%%%%%%%%%%%%%%%%%%%%%%%%%%%%%%%%%%%%%%%%
% Statement
\section*{Competing financial interests}
The authors declare that they have no competing financial interests.

%%%%%%%%%%%%%%%%%%%%%%%%%%%%%%%%%%%%%%%%%%%%%%%%%%%%%%%%%%%%%%%%%%%%%%%%%
%%%%%%%%%%%%%%%%%%%%%%%%%%%%%%%%%%%%%%%%%%%%%%%%%%%%%%%%%%%%%%%%%%%%%%%%%
% References
%\section*{References}

\hyphenation{Post-Script Sprin-ger}

%%%%%%%%%%%%%%%%%%%%%%%%%%%%%%%%%%%%%%%%%%%%%%%%%%%%%%%%%%%%%%%%%%%%%%%%%
%%%%%%%%%%%%%%%%%%%%%%%%%%%%%%%%%%%%%%%%%%%%%%%%%%%%%%%%%%%%%%%%%%%%%%%%%
% Figure legends
%\newpage
%\section*{Figure legends}

\begin{thebibliography}{10}
\expandafter\ifx\csname url\endcsname\relax
  \def\url#1{\texttt{#1}}\fi
\expandafter\ifx\csname urlprefix\endcsname\relax\def\urlprefix{URL }\fi
\providecommand{\bibinfo}[2]{#2}
\providecommand{\eprint}[2][]{\url{#2}}

\bibitem{PaulScience2001}
\bibinfo{author}{Paul, P.~M.} \emph{et~al.}
\newblock \bibinfo{title}{Observation of a {T}rain of {A}ttosecond {P}ulses
  from {H}igh {H}armonic {G}eneration}.
\newblock \emph{\bibinfo{journal}{Science}}
  \textbf{\bibinfo{volume}{\textbf{292}}}, \bibinfo{pages}{1689}
  (\bibinfo{year}{2001}).

\bibitem{DrescherScience2001}
\bibinfo{author}{Drescher, M.} \emph{et~al.}
\newblock \bibinfo{title}{X-{R}ay {P}ulses {A}pproaching the {A}ttosecond
  {F}rontier}.
\newblock \emph{\bibinfo{journal}{Science}} \textbf{\bibinfo{volume}{291}},
  \bibinfo{pages}{1923} (\bibinfo{year}{2001}).

\bibitem{KienbergerScience2002}
\bibinfo{author}{Kienberger, R.} \emph{et~al.}
\newblock \bibinfo{title}{Steering {A}ttosecond {E}lectron {W}ave {P}ackets
  with {L}ight}.
\newblock \emph{\bibinfo{journal}{Science}}
  \textbf{\bibinfo{volume}{\textbf{297}}}, \bibinfo{pages}{1144}
  (\bibinfo{year}{2002}).

\bibitem{KienbergerNature2004}
\bibinfo{author}{Kienberger, R.} \emph{et~al.}
\newblock \bibinfo{title}{Atomic transient recorder}.
\newblock \emph{\bibinfo{journal}{Nature}} \textbf{\bibinfo{volume}{427}},
  \bibinfo{pages}{817} (\bibinfo{year}{2004}).

\bibitem{GoulielmakisScience2004}
\bibinfo{author}{Goulielmakis, E.} \emph{et~al.}
\newblock \bibinfo{title}{Direct {M}easurement of {L}ight {W}aves}.
\newblock \emph{\bibinfo{journal}{Science}} \textbf{\bibinfo{volume}{305}},
  \bibinfo{pages}{1267} (\bibinfo{year}{2004}).

\bibitem{JohnssonPRL2005}
\bibinfo{author}{Johnsson, P.} \emph{et~al.}
\newblock \bibinfo{title}{Attosecond {E}lectron {W}ave {P}acket {D}ynamics in
  {S}trong {L}aser {F}ields}.
\newblock \emph{\bibinfo{journal}{Phys. {R}ev. {L}ett.}}
  \textbf{\bibinfo{volume}{95}}, \bibinfo{pages}{013001}
  (\bibinfo{year}{2005}).

\bibitem{RemetterNP2006}
\bibinfo{author}{Remetter, T.} \emph{et~al.}
\newblock \bibinfo{title}{Attosecond {E}lectron {W}ave {P}acket
  {I}nterferometry}.
\newblock \emph{\bibinfo{journal}{Nature {P}hys.}}
  \textbf{\bibinfo{volume}{2}}, \bibinfo{pages}{323} (\bibinfo{year}{2006}).

\bibitem{SchererJCP1991}
\bibinfo{author}{Scherer, N.~F.} \emph{et~al.}
\newblock \bibinfo{title}{Fluorescence-detected wave packet interferometry:
  {T}ime resolved molecular spectroscopy with sequences of femtosecond
  phase-locked pulses}.
\newblock \emph{\bibinfo{journal}{J. {C}hem. {P}hys.}}
  \textbf{\bibinfo{volume}{95}}, \bibinfo{pages}{1487} (\bibinfo{year}{1991}).

\bibitem{NoordamPRA1992}
\bibinfo{author}{Noordam, L.~D.}, \bibinfo{author}{Duncan, D.~I.} \&
  \bibinfo{author}{Gallagher, T.~F.}
\newblock \bibinfo{title}{Ramsey fringes in atomic {R}ydberg wave packets}.
\newblock \emph{\bibinfo{journal}{Phys. {R}ev. {A}}}
  \textbf{\bibinfo{volume}{45}}, \bibinfo{pages}{4734} (\bibinfo{year}{1992}).

\bibitem{KulanderAILF1992}
\bibinfo{author}{Kulander, K.~C.}, \bibinfo{author}{Schafer, K.~J.} \&
  \bibinfo{author}{Krause, J.~L.}
\newblock \bibinfo{title}{Time-{D}ependent {S}tudies of {M}ultiphoton
  {P}rocesses}.
\newblock In \emph{\bibinfo{booktitle}{Atoms in {I}ntense {L}aser {F}ields}}
  (\bibinfo{publisher}{Academic Press}, \bibinfo{address}{San Diego},
  \bibinfo{year}{1992}).

\bibitem{KrausePRL1992}
\bibinfo{author}{Krause, J.~L.}, \bibinfo{author}{Schafer, K.~J.} \&
  \bibinfo{author}{Kulander, K.~C.}
\newblock \bibinfo{title}{High-order harmonic generation from atoms and ions in
  the high intensity regime}.
\newblock \emph{\bibinfo{journal}{Phys. {R}ev. {L}ett.}}
  \textbf{\bibinfo{volume}{\textbf{68}}}, \bibinfo{pages}{3535}
  (\bibinfo{year}{1992}).

\bibitem{SolaNP2006}
\bibinfo{author}{Sola, I.~J.} \emph{et~al.}
\newblock \bibinfo{title}{Controlling attosecond electron dynamics by
  phase-stabilized polarization gating}.
\newblock \emph{\bibinfo{journal}{Nature {P}hys.}}
  \textbf{\bibinfo{volume}{2}}, \bibinfo{pages}{319} (\bibinfo{year}{2006}).

\bibitem{DrescherNature2002}
\bibinfo{author}{Drescher, M.} \emph{et~al.}
\newblock \bibinfo{title}{Time-resolved atomic inner-shell spectroscopy}.
\newblock \emph{\bibinfo{journal}{Nature}} \textbf{\bibinfo{volume}{419}},
  \bibinfo{pages}{803} (\bibinfo{year}{2002}).

\bibitem{LopezMartensPRL2005}
\bibinfo{author}{{L\'opez-Martens}, R.} \emph{et~al.}
\newblock \bibinfo{title}{Amplitude and {P}hase {C}ontrol of {A}ttosecond
  {L}ight {P}ulses}.
\newblock \emph{\bibinfo{journal}{Phys. {R}ev. {L}ett.}}
  \textbf{\bibinfo{volume}{94}}, \bibinfo{pages}{033001}
  (\bibinfo{year}{2005}).

\bibitem{MullerAPB2002}
\bibinfo{author}{Muller, H.~G.}
\newblock \bibinfo{title}{Reconstruction of attosecond harmonic beating by
  interference of two-photon transitions}.
\newblock \emph{\bibinfo{journal}{Appl. {P}hys.~{B}}}
  \textbf{\bibinfo{volume}{\textbf{74}}}, \bibinfo{pages}{17}
  (\bibinfo{year}{2002}).

\bibitem{BucksbaumJOSAB1987}
\bibinfo{author}{Bucksbaum, P.~H.}, \bibinfo{author}{Freeman, R.~R.},
  \bibinfo{author}{Bashkansky, M.} \& \bibinfo{author}{McIlratht, T.~J.}
\newblock \bibinfo{title}{Role of the ponderomotive potential in
  above-threshold ionization}.
\newblock \emph{\bibinfo{journal}{J. {O}pt. {S}oc. {A}m.~{B}}}
  \textbf{\bibinfo{volume}{\textbf{4}}}, \bibinfo{pages}{760}
  (\bibinfo{year}{1987}).

\bibitem{EppinkRSI1997}
\bibinfo{author}{Eppink, A. T. J.~B.} \& \bibinfo{author}{Parker, D.~H.}
\newblock \bibinfo{title}{Velocity map imaging of ions and electrons using
  electrostatic lenses: {A}pplication in photoelectron and photofragment ion
  imaging of molecular oxygen}.
\newblock \emph{\bibinfo{journal}{Rev. {S}ci. {I}nstrum.}}
  \textbf{\bibinfo{volume}{68}}, \bibinfo{pages}{3477} (\bibinfo{year}{1997}).

\bibitem{VrakkingRSI2001}
\bibinfo{author}{Vrakking, M. J.~J.}
\newblock \bibinfo{title}{An iterative procedure for the inversion of
  two-dimensional ion/photoelectron imaging experiments}.
\newblock \emph{\bibinfo{journal}{Rev. {S}ci. {I}nstr.}}
  \textbf{\bibinfo{volume}{72}}, \bibinfo{pages}{4084} (\bibinfo{year}{2001}).

\bibitem{WalkerPRL1994}
\bibinfo{author}{Walker, B.}
\newblock \bibinfo{title}{Precision {M}easurement of {S}trong {F}ield {D}ouble
  {I}onization of {H}elium}.
\newblock \emph{\bibinfo{journal}{Phys. {R}ev. {L}ett.}}
  \textbf{\bibinfo{volume}{73}}, \bibinfo{pages}{1227} (\bibinfo{year}{1994}).

\bibitem{GaardePRL2000}
\bibinfo{author}{Gaarde, M.~B.} \emph{et~al.}
\newblock \bibinfo{title}{Strong species dependence of high order photoelectron
  production in alkali metal atoms}.
\newblock \emph{\bibinfo{journal}{Phys. {R}ev. {L}ett.}}
  \textbf{\bibinfo{volume}{84}}, \bibinfo{pages}{2822--2825}
  (\bibinfo{year}{2000}).

\bibitem{KulanderCPC1991}
\bibinfo{author}{Kulander, K.} \& \bibinfo{author}{Rescigno, T.}
\newblock \bibinfo{title}{Effective potentials for time-dependent calculations
  of multiphoton processes in atoms}.
\newblock \emph{\bibinfo{journal}{Computer {P}hysics {C}ommunications}}
  \textbf{\bibinfo{volume}{63}}, \bibinfo{pages}{523} (\bibinfo{year}{1991}).

\bibitem{SchaferPRA1990}
\bibinfo{author}{Schafer, K.~J.} \& \bibinfo{author}{Kulander, K.~C.}
\newblock \bibinfo{title}{The energy analysis of time-dependent wave functions:
  {A}pplication to above threshold ionization}.
\newblock \emph{\bibinfo{journal}{Phys. {R}ev. {A}}}
  \textbf{\bibinfo{volume}{42}}, \bibinfo{pages}{5794} (\bibinfo{year}{1990}).

\end{thebibliography}
\end{document}